\documentclass[showpacs,preprintnumbers,amsmath,amssymb,twocolumn]{revtex4}%

\usepackage{graphicx}% Include figure files
\usepackage{dcolumn}% Align table columns on decimal point
\usepackage{bm}% bold math
\usepackage{color}

\begin{document}

\title{Rotating Regular Black Hole Solution}

\author{Bobir Toshmatov$^{1}$}
\email{b.a.toshmatov@gmail.com}

\author{Bobomurat Ahmedov$^{1,2}$}
\email{ahmedov@astrin.uz}

\author{Ahmadjon Abdujabbarov$^{1,2,3}$}
\email{ahmadjon@astrin.uz}

\author{Zden\v{e}k Stuchl\'{i}k$^{3}$}
\email{zdenek.stuchlik@fpf.slu.cz}

\affiliation{%
$^{1}$ Institute of Nuclear Physics, Ulughbek, Tashkent 100214,
Uzbekistan\\
$^{2}$ Ulugh Beg Astronomical Institute, Astronomicheskaya 33,
Tashkent 100052, Uzbekistan\\
$^{3}$ Institute of Physics, Faculty of Philosophy \& Science, Silesian University in Opava,
Bezru\v{c}ovo n\'{a}m\v{e}st\'{i} 13, CZ-74601 Opava, Czech Republic}

\begin{abstract}
Based on the Newman-Janis algorithm the Ay\'{o}n-Beato-Garc\'{i}a
spacetime metric~\cite{Eloy1} of the regular spherically
symmetric, static and charged black hole has been converted into
rotational form. It is shown that the derived solution for
rotating regular black hole is regular and the critical value of
the electric charge $Q$ for which two horizons merge into one
sufficiently decreases in the presence of nonvanishing angular momentum $a$ of the black hole.
\end{abstract}

\maketitle

\section{Introduction}

It is well known that exact solutions of the Einstein equations
have one of the "mysterious" properties of the black hole which is
called singularity. Singularity has been considered as one of
defects of the general relativity because explanation of
singularity can not be made by the general relativity itself. So
called regular black hole solutions \cite{Eloy1}, \cite{Eloy},
\cite{Eloy2} can be created in order to eliminate singularity from
the spacetime metric.  

One can distinguish 
several types 
of the regular black 
hole
solutions: 
solutions which 
are continuous 
everywhere
throughout spacetime;
solutions with 
two simple regions and 
boundary surfaces 
joining these regions;  
solutions with
separated two regions: 
surface layer,
thin shell, 
joining the two 
regions, etc. 
The 
regular black hole solution 
obtained in the 
Refs.~\cite{Eloy1,Eloy} 
belongs to
the first type 
of the regular 
black holes 
mentioned above.

There exist two types of singularities: the coordinate one
(event horizon) and the curvature singularity where 
the curvature of the manifold is becoming infinite. In
the case of coordinate singularity, $g_{rr}$ component of the
metric tensor goes to infinity. One can eliminate coordinate
singularity by making transformation to the more fortunate
coordinate system. Usually by transforming coordinates from the
Boyer-Lindquist coordinates to the Eddington-Finkelstein ones one
can remove coordinate singularity from the spacetime metric.
Eddington-Finkelstein coordinates are based on the freely falling
photons. On the other hand, in the curvature singularity, the
Riemann tensor components of the spacetime metric diverge. It is
impossible to eliminate curvature singularity from the spacetime
metric by coordinate transformations.

In the papers \cite{Eloy1}, \cite{Eloy} and \cite{Eloy2}, the new
regular black hole solutions of the Einstein equations has been
found taking into account the coupling to non-linear
electrodynamic field. Afterwards, this solution has been called
Ay\'{o}n-Beato-Garc\'{i}a regular black hole solution. Recently
another regular black hole solution ~\cite{neves} has been
considered by introducing a new mass function generalizing the
commonly used Bardeen and Hayward mass functions and including the
cosmological constant.

The Kerr spacetime metric can be derived from the Schwarzschild
one by using the Newman-Janis algorithm \cite{Drake},
\cite{Bambi}. The derivation of the Kerr spacetime metric from
Schwarzschild one has been given in several works \cite{Drake},
\cite{Drake2} and \cite{Julio}. Moreover, in the papers
\cite{Drake} and \cite{Julio},  the Kerr-Newman solution has been
derived from the Reissner-Nordstr\"{o}m spacetime metric. The
Newman-Janis algorithm has been used to derive the radiating
Kerr-Newman black hole in $f( R)$ gravity~\cite{naresh1}. The
exact nonstatic charged BTZ-like solutions, in (N+1)-dimensional
Einstein gravity have been found in~\cite{naresh2} in the presence
of negative cosmological constant. The Lovelock gravity in the
critical spacetime dimension has been studied in
Ref.~\cite{naresh3}.

In order to convert static, spherically symmetric black hole
spacetime metric into rotational one (if this spacetime metric is
given in the Boyer-Lindquist coordinates ($t$, $r$, $\theta$,
$\phi$)) one has to proceed the following five steps of the
Newman-Janis algorithm: i) transition from the Boyer-Lindquist
coordinates into the advanced Eddington-Filkenstein ones ($u$,
$r$, $\theta$, $\phi$) has to be performed; ii) a null tetrad
($\textbf{l}$, $\textbf{n}$, $\textbf{m}$ and $\textbf{\={m}}$)
(Newman-Penrose tetrad) for produced metric have to be found; iii)
a complex coordinate transformations has to be applied; iv)
reverse coordinate transformations into the Boyer-Lindquist ones
have to be done; v) finally, unknown terms of the transformations
have to be found based on the reality condition.

Here we convert the static, spherically symmetric
Ay\'{o}n-Beato-Garc\'{i}a regular black hole spacetime
\cite{Eloy1}, \cite{Eloy}, \cite{Eloy2} into the rotational one by
using the Newman-Janis algorithm \cite{Drake}, \cite{Bambi} and
study some of its basic properties.

%Throughout the paper, we use a space-like signature $(-,+,+,+)$
%and a system of units in which $c = 1 = G$, $c$ being the velocity
%of light in vacuum and $G$ the gravitational constant. Greek
%indices are taken to run from 0 to 3 and Latin indices from 1 to
%3; covariant derivatives are denoted with a semi-colon and partial
%derivatives with a comma.

\section{Newman-Janis algorithm to get rotating regular black hole solution }

In this section we describe the Newman-Janis algorithm which is used for
converting the spherically symmetric static black hole spacetime
metric into rotational one. The Ay\'{o}n-Beato-Garc\'{i}a
spacetime metric of the regular spherically symmetric black hole
is given as \cite{Garcia}
\begin{eqnarray}\label{0}
ds^2=-f(r)dt^2+\frac{dr^2}{f(r)}+r^2d\theta^2+r^2\sin^2{\theta}d\phi^2,
\end{eqnarray}
where the lapse function $f(r)$ reads
\begin{eqnarray}\label{01}
f(r)=1-\frac{2M r^2}{(r^2+Q^2)^{3/2}}+\frac{Q^2r^2}{(r^2+Q^2)^2}\ ,
\end{eqnarray}
$M$ and $Q$ are the total mass and electric charge of the black
hole, respectively. The spacetime metric (\ref{0}) is the solution
of the field equations within general relativity, where the
nonlinear electrodynamic field satisfying the weak energy
condition is considered as a source.
As can be seen from the lapse function (\ref{01}) the spacetime
metric (\ref{0}) has only the coordinate singularity. This is why
in order to remove this singularity one has to write the spacetime
metric (\ref{0}) in the advanced Eddington-Finkelstein
coordinates.  To do this we make following transformation for
incoming photon (or ray):
\begin{eqnarray}\label{02}
v=t-r^\ast,
\end{eqnarray}
and for outgoing photon (or ray)
\begin{eqnarray}\label{03}
u=t+r^\ast,
\end{eqnarray}
where
\begin{eqnarray}\label{04}
r^\ast=\int\frac{dr}{f(r)}.
\end{eqnarray}
Hereafter, we consider only the outgoing photon (\ref{03}) case.
Then the spacetime metric (\ref{0}) in the advanced
Eddington-Finkelstein coordinates takes the form
\begin{eqnarray}\label{05}
ds^2=-f(r)du^2-2dudr+r^2d\theta^2+r^2\sin^2{\theta}d\phi^2.
\end{eqnarray}
The Newman-Penrose tetrad consists of four isotropic vectors
$\textbf{l}$, $\textbf{n}$, $\textbf{m}$ and $\textbf{\={m}}$.
$\textbf{l}$ and $\textbf{n}$ are real vectors, $\textbf{m}$ and
$\textbf{\={m}}$ are mutually complex conjugate vectors
\cite{Julio}.

Newman-Penrose tetrads satisfy orthogonality condition:
\begin{eqnarray}\label{1}
l^\mu \cdot m_{\mu}=l^\mu \cdot \bar{m}_{\mu}=n^\mu \cdot
m_{\mu}=n^\mu \cdot \bar{m}_{\mu}=0,
\end{eqnarray}
and also isotropic condition:
\begin{eqnarray}\label{2}
l^\mu \cdot l_{\mu}=n^\mu \cdot n_{\mu}=m^\mu \cdot
m_{\mu}=\bar{m}^\mu \cdot \bar{m}_{\mu}=0.
\end{eqnarray}
Moreover, the basis vectors usually impose the following
normalization condition:
\begin{eqnarray}\label{3}
l^\mu \cdot n_{\mu}=1, & m^\mu \cdot \bar{m}_{\mu}=-1,
\end{eqnarray}
where $\bar{m}^{\mu}$ is the complex conjugate of $m^{\mu}$.

The contravariant components of the metric tensor of the spacetime
metric (\ref{05}) are
\begin{eqnarray} \label{contr}
g^{\mu\nu} =      \left(
\begin{array}{cccc}
0 & -1 & 0 &0 \\
-1 & f(r) & 0 & 0 \\
0 & 0 & 1/r^2 & 0 \\
0 & 0 & 0 & 1/r^2\sin^2\theta \\
\end{array} \right).
\end{eqnarray}

We can rewrite (\ref{contr}) with the help of Newman-Penrose
tetrad as
\begin{eqnarray}\label{4}
g^{\mu\nu}=-l^\mu \cdot n^{\nu}-l^\nu \cdot n^{\mu}+m^\mu \cdot
\bar{m}^{\nu}+m^\nu \cdot \bar{m}^{\mu},
\end{eqnarray}
where the components of the null tetrad vectors are
\begin{eqnarray}\label{5}
l^\mu=[0,1,0,0], n^\mu=[1,-\frac{1}{2}f(r),0,0],\nonumber\\
m^\mu=\frac{1}{\sqrt{2}r}[0,0,1,\frac{i}{\sin\theta}],
\bar{m}^{\mu}=\frac{1}{\sqrt{2}r}[0,0,1,-\frac{i}{\sin\theta}].
\end{eqnarray}

As the next step we make the following complex coordinate
transformations:
\begin{eqnarray}\label{6}
\tilde{r}=r+ia\cos\theta, & \tilde{u}=u-ia\cos\theta,\nonumber\\
\tilde{\theta}=\theta, & \tilde{\phi}=\phi.
\end{eqnarray}
As a result of these transformations the components of the null
tetrad vectors take the form \cite{Drake2}
\begin{eqnarray}\label{7}
\tilde{l}^\mu=[0,1,0,0],  \tilde{n}^\mu=[1,-\frac{1}{2}\tilde{f}(r),0,0],\nonumber\\
\tilde{m}^\mu=\frac{1}{\sqrt{2}(r+ia\cos\theta)}[ia\sin\theta,-ia\sin\theta,1,\frac{i}{\sin\theta}],\nonumber\\
\tilde{\bar{m}}^\mu=\frac{1}{\sqrt{2}(r-ia\cos\theta)}[-ia\sin\theta,ia\sin\theta,1,-\frac{i}{\sin\theta}],
\end{eqnarray}
where the function
\begin{eqnarray}\label{8}
\tilde{f}(r)=1-\frac{2M r
\sqrt{\Sigma}}{(\Sigma+Q^2)^{3/2}}+\frac{Q^2\Sigma}{(\Sigma+Q^2)^2}
\end{eqnarray}
is the new form of the lapse function (\ref{6}) and \\
$\Sigma=r^2+a^2\cos^2\theta$.

Then the metric tensor $g^{\mu\nu}$ takes new $\tilde{g}^{\mu\nu}$
form
\begin{eqnarray}\label{9}
\tilde{g}^{\mu\nu}=-\tilde{l}^\mu \cdot
\tilde{n}^{\nu}-\tilde{l}^\nu \cdot \tilde{n}^{\mu}+\tilde{m}^\mu
\cdot \tilde{\bar{m}}^{\nu}+\tilde{m}^\nu \cdot
\tilde{\bar{m}}^{\mu},
\end{eqnarray}
or
\begin{eqnarray} \label{contr1}
\tilde{g}^{\mu\nu} =      \left(
\begin{array}{cccc}
\frac{a^2\sin^2\theta}{\Sigma} & -1-\frac{a^2\sin^2\theta}{\Sigma} & 0 &\frac{a}{\Sigma} \\
-1-\frac{a^2\sin^2\theta}{\Sigma} & \tilde{f}(r)+\frac{a^2\sin^2\theta}{\Sigma} & 0 & -\frac{a}{\Sigma} \\
0 & 0 & \frac{1}{\Sigma} & 0 \\
\frac{a}{\Sigma} & -\frac{a}{\Sigma} & 0 & \frac{1}{\Sigma\sin^2\theta} \\
\end{array} \right).
\end{eqnarray}
The covariant components of the metric tensor (\ref{contr1}) are
\begin{widetext}
\begin{eqnarray} \label{contr2}
\tilde{g}_{\mu\nu} =      \left(
\begin{array}{cccc}
-\tilde{f}(r) & -1 & 0 &a(\tilde{f}(r)-1)\sin^2\theta \\
-1 & 0 & 0 & a\sin^2\theta \\
0 & 0 & \Sigma & 0 \\
a(\tilde{f}(r)-1)\sin^2\theta & a\sin^2\theta & 0 & \sin^2\theta[\Sigma-a^2(\tilde{f}(r)-2)\sin^2\theta] \\
\end{array} \right),
\end{eqnarray}
\end{widetext}
and the spacetime element can be written as
\begin{eqnarray}\label{10}
d\tilde{s}^2=g_{uu}du^2+2g_{ur}dudr+2g_{u\phi}dud\phi+\nonumber\\
2g_{r\phi}drd\phi+g_{\theta\theta}d\theta^2+g_{\phi\phi}d\phi^2.
\end{eqnarray}
At the following step we turn back into the Boyer-Lindquist
coordinates using, the following transformations:
\begin{eqnarray}\label{11}
du=dt+\lambda(r)dr, & d\phi=d\phi+\chi(r)dr,
\end{eqnarray}
where functions $\lambda(r)$ and $\chi(r)$ are chosen for
eliminating $g_{tr}$ and $g_{r\phi}$ terms. Putting (\ref{11})
into (\ref{10}) and collecting terms which are corresponding to
$g_{tr}$ and $g_{r\phi}$ ones, then equalizing produced expression
to zero, one can get two equations for two unknown functions
$\lambda(r)$, $\chi(r)$. Solving these equations simultaneously
one can find expressions for $\lambda(r)$ and $\chi(r)$ in the
following way
\begin{eqnarray}\label{12}
\lambda(r)=-\frac{\Sigma+a^2\sin^2\theta}{\Sigma\tilde{f}(r)+a^2\sin^2\theta},\nonumber\\
\chi(r)=-\frac{a}{\Sigma\tilde{f}(r)+a^2\sin^2\theta}.
\end{eqnarray}
Finally, the spacetime metric can be expressed in the
Boyer-Lindquist coordinates as
%
%\begin{widetext}
%\begin{eqnarray} \label{contr3}
%\tilde{g}_{\mu\nu} =      \left(
%\begin{array}{cccc}
%-\tilde{f}(r) & 0 & 0 & a(\tilde{f}(r)-1)\sin^2\theta \\
%0 & \frac{\Sigma}{\Sigma\tilde{f}(r)+a^2\sin^2\theta} & 0 & 0 \\
%0 & 0 & \Sigma & 0 \\
%a(\tilde{f}(r)-1)\sin^2\theta & 0 & 0 & \sin^2\theta[\Sigma-a^2(\tilde{f}(r)-2)\sin^2\theta] \\
%\end{array} \right),
%\end{eqnarray}
%\end{widetext}
%
as
\begin{eqnarray}\label{13}
d\tilde{s}^2&=&-\tilde{f}(r)dt^2+\frac{\Sigma}{\Sigma\tilde{f}(r)+a^2\sin^2\theta}dr^2\nonumber\\&&-2a\sin^2\theta(1-\tilde{f}(r))d\phi dt+\Sigma d\theta^2\nonumber\\
&& +\sin^2\theta[\Sigma-a^2(\tilde{f}(r)-2)\sin^2\theta]d\phi^2,
\end{eqnarray}
where $\tilde{f}(r)$ is given by equation (\ref{8}).

%=1-{2M r
%\sqrt{\Sigma}}/{(\Sigma+Q^2)^{3/2}}+{Q^2\Sigma}/{(\Sigma+Q^2)^2}$.

If we consider the black hole as non-charged one ($Q=0$), the
lapse function (\ref{01}) takes the same form of one of the
Schwarzschild spacetime metric as well as new spacetime metric
(\ref{13}) converts into the Kerr one, namely
\begin{eqnarray}\label{14}
ds^2=-(1-\frac{2M r}{\Sigma})dt^2+\frac{\Sigma}{\Delta}dr^2
-2\frac{2M r}{\Sigma}a\sin^2\theta d\phi dt\nonumber\\ +\Sigma
d\theta^2 +(r^2+a^2+\frac{2Ma^2r\sin^2\theta}{\Sigma})\sin^2\theta
d\phi^2,\nonumber\\
\end{eqnarray}
where $\Delta=r^2+a^2-2M r$, $\sum=r^2+a^2\cos^2\theta$.

In order to investigate properties of the spacetime metric
(\ref{13}) we here study the behavior of the $g_{rr}$ and $g_{tt}$
components of the spacetime metric (\ref{13}).

\section{Properties Of Rotating Regular Black Hole Solution}

\begin{figure*}[t!.]
\begin{center}
\includegraphics[width=0.32\linewidth]{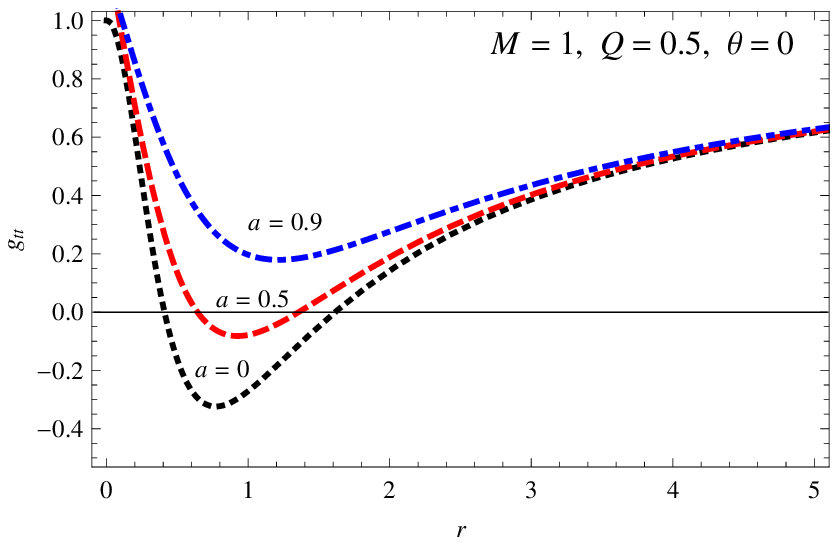}
\includegraphics[width=0.32\linewidth]{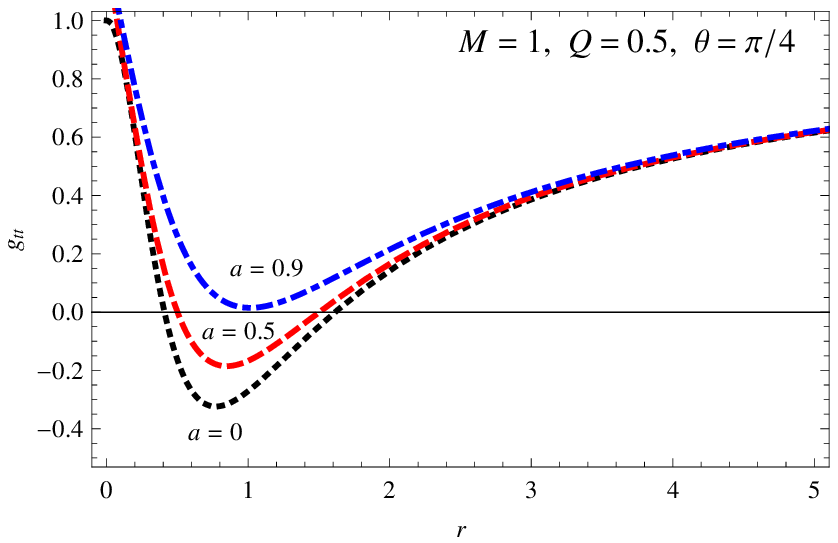}
\includegraphics[width=0.32\linewidth]{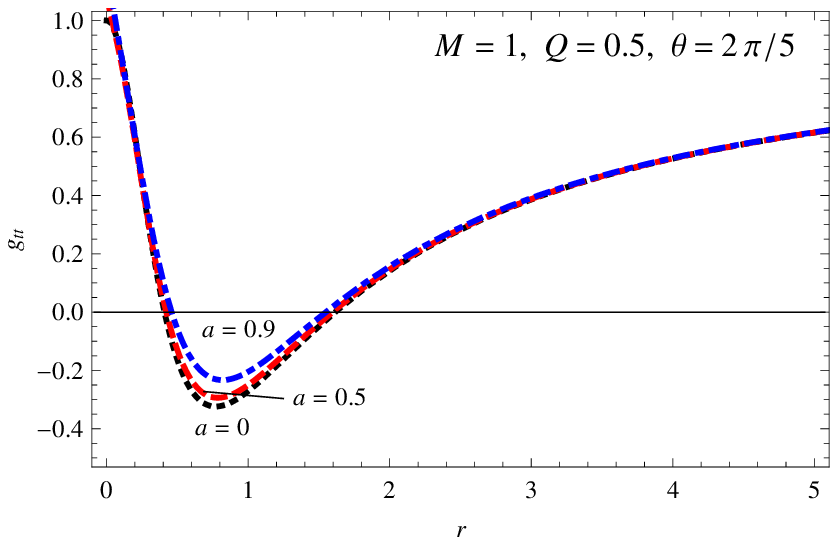}
\end{center}
\caption{\label{plot2} The dependence of the $g_{tt}$ component of
the metric tensor from the radial coordinate $r$ for the typical
values of the rotation parameter $a$.}
\end{figure*}
\begin{figure*}[t!.]
\begin{center}
\includegraphics[width=0.32\linewidth]{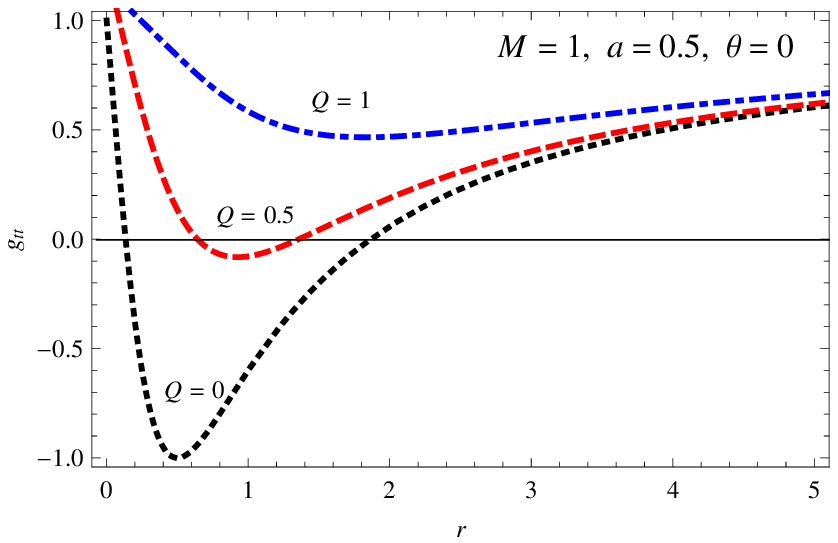}
\includegraphics[width=0.32\linewidth]{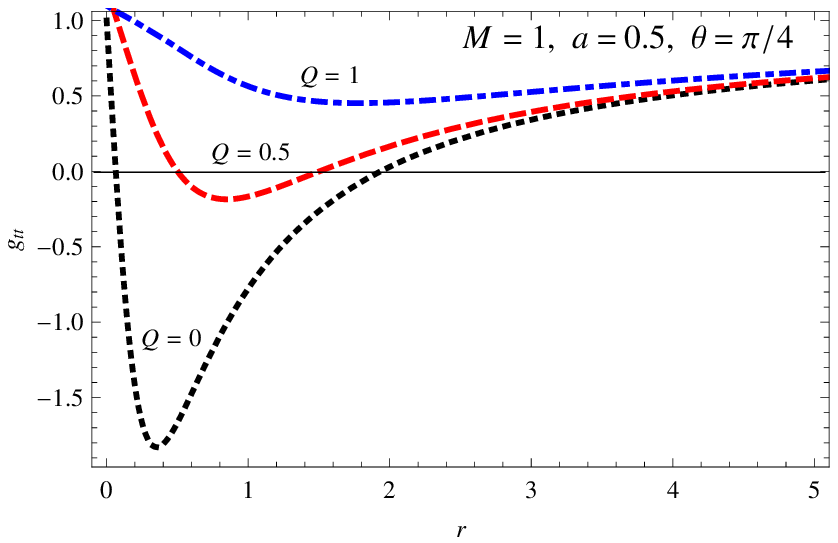}
\includegraphics[width=0.32\linewidth]{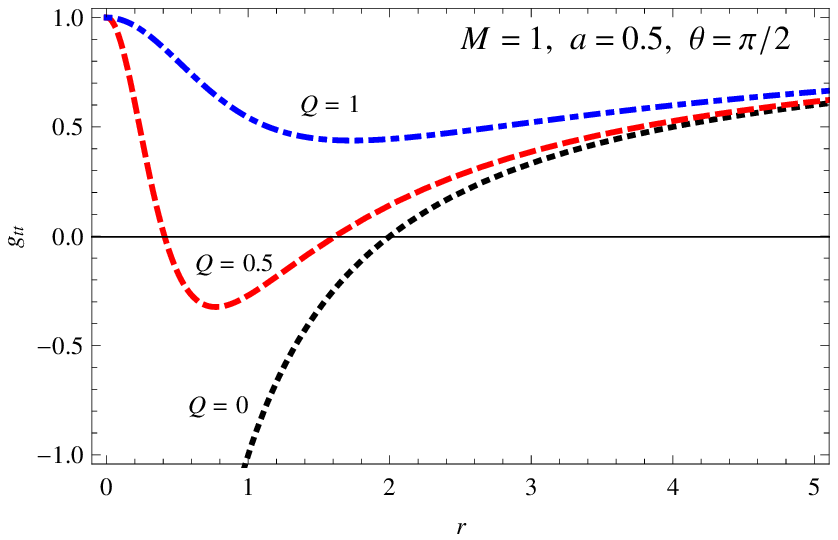}
\end{center}
\caption{\label{plot3} The dependence of the $g_{tt}$ component of
the metric tensor from the radial coordinate $r$ for the typical
values of the electric charge $Q$.}
\end{figure*}
\begin{figure*}[t!.]
\begin{center}
\includegraphics[width=0.32\linewidth]{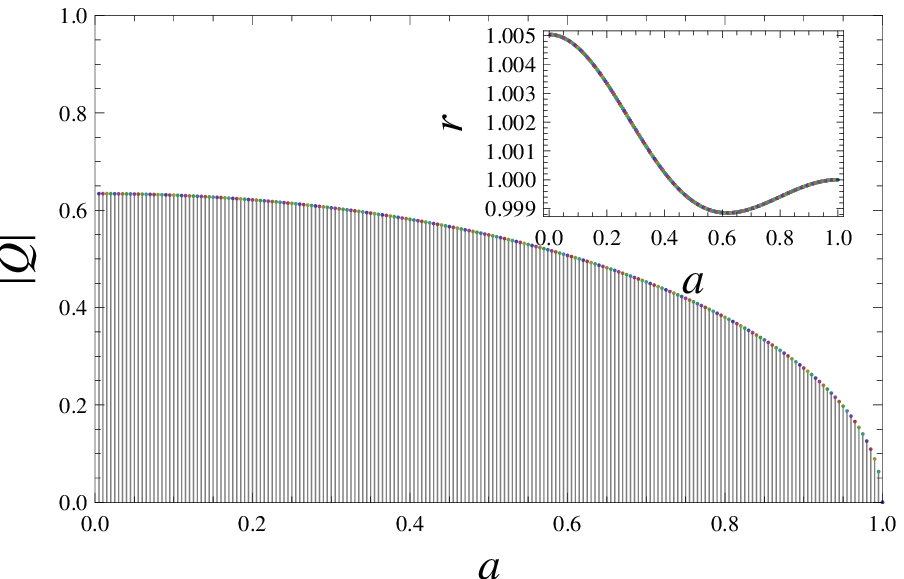}
\includegraphics[width=0.32\linewidth]{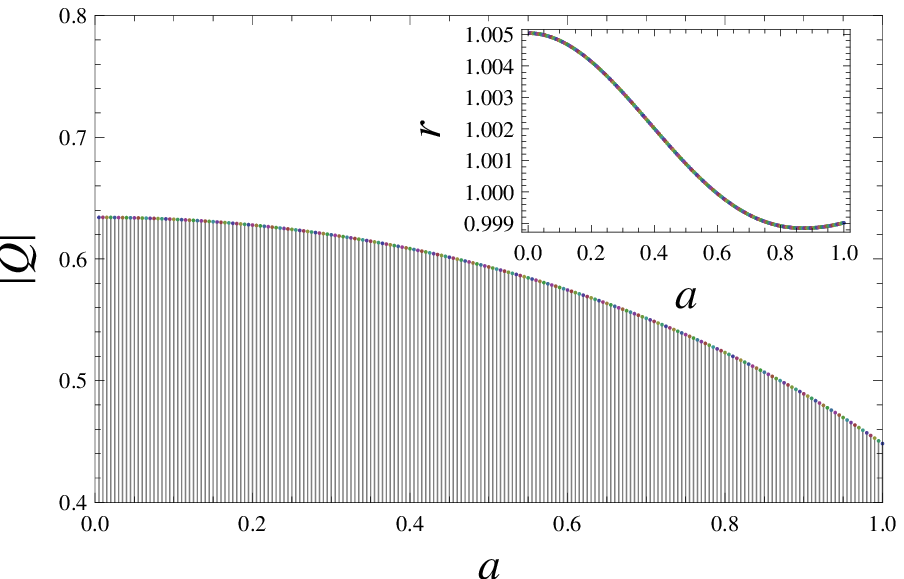}
\includegraphics[width=0.32\linewidth]{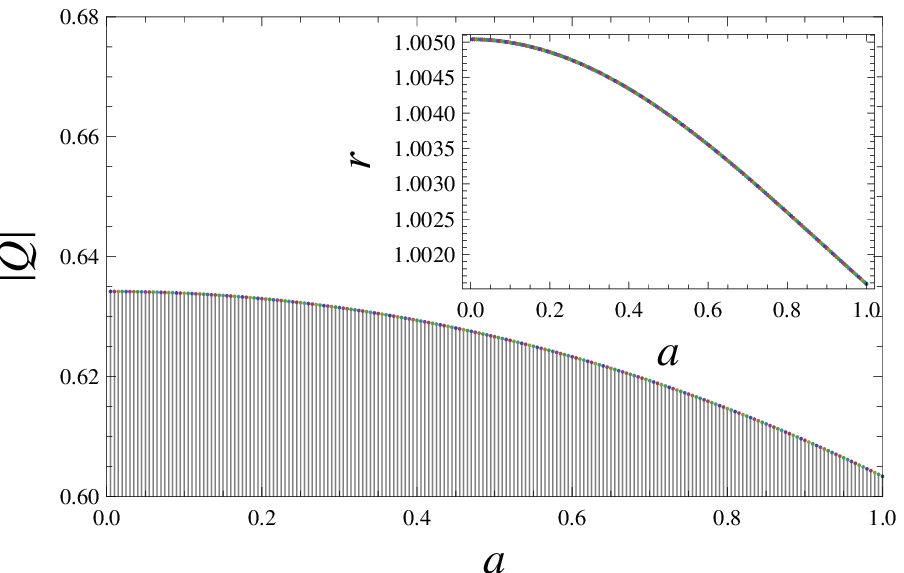}
\end{center}
\caption{\label{fig_ahmadjon} The dependence of the critical value
of the electric charge $Q$ and radius of the horizon $r$ from the
rotation parameter $a$ for the different values of $\theta$:
$\theta=0$, $\theta=\pi/4$ and $\theta=2\pi/5$ (from the left to
right, respectively).}
\end{figure*}
\begin{figure*}[t!.]
\begin{center}
\includegraphics[width=0.32\linewidth]{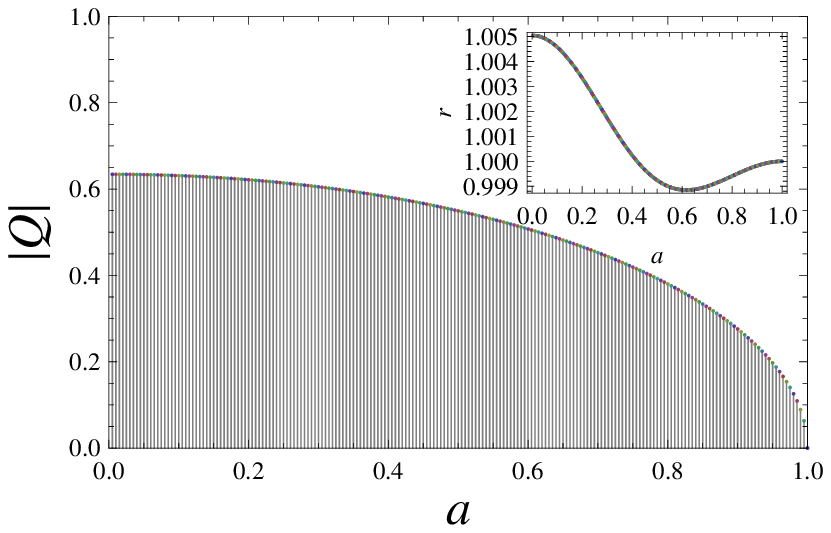}
\includegraphics[width=0.32\linewidth]{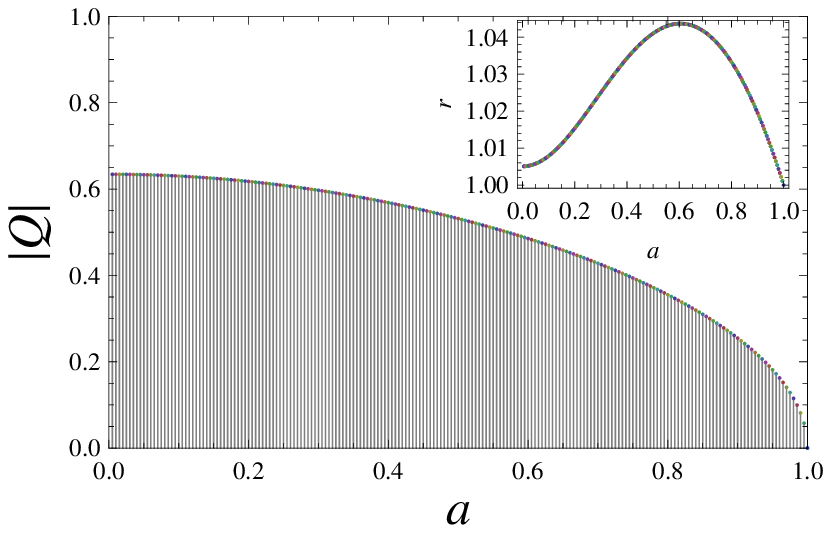}
\includegraphics[width=0.32\linewidth]{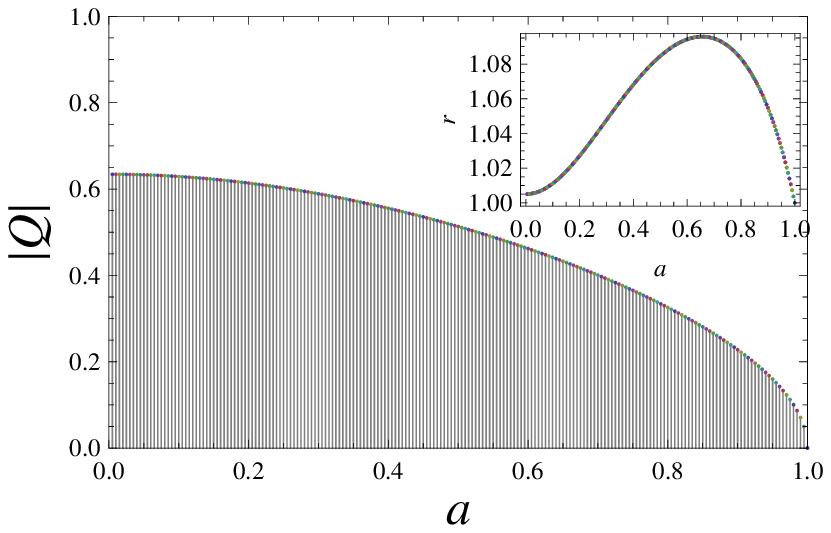}
\end{center}
\caption{\label{fig_ahmadjon2} The dependence of the critical
value of the electric charge $Q$ and radius of the event horizon
$r$ from the rotation parameter $a$ for the different values of
$\theta$: $\theta=0$, $\theta=\pi/4$ and $\theta=\pi/2$ (from the
left to right, respectively).}
\end{figure*}

{Now we will analyze the  static limit and event
horizon defined by the conditions  $g_{tt}=0$ and $1/g_{rr}=0$,
respectively.}

Obtained new spacetime metric (\ref{13}) is also regular
everywhere. From the Fig.~\ref{plot2} and Fig.~\ref{plot3} one can
easily see that for some set of values of the rotation parameter
$a$ and electric charge $Q$ the solution (\ref{13}) has coordinate
singularity (event horizon).

The radial dependence of the function $\tilde{f}(r)$ presented in
Fig.~\ref{plot2} and Fig.~\ref{plot3} show that with increase of
the value of the rotation parameter $a$ and charge $Q$ possibility
of existence of the horizon decreases. In the equatorial plane
($\theta=\pi/2$) the dependence of the function $\tilde{f}(r)$ on
the rotation parameter $a$ vanishes and the existence of the
horizon depends only on the value of the charge $Q$.

There is critical value of the charge $Q$ for which two surface described by the solutions of the condition  $g_{tt}=0$
merge into one. In order to find the critical value of $Q$ the
lapse function $g_{tt}(r,a,\theta,Q)$ must satisfy the couple of
conditions:
\begin{eqnarray}\label{001}
g_{tt}(r,a,\theta,Q)=0, & \partial_{r}g_{tt}(r,a,\theta,Q)=0.
\end{eqnarray}
Since the lapse function $g_{tt}$ is the function of four
quantities $r,a,\theta$ and $Q$, solving the equations (\ref{001})
with respect to $r$ and $Q$ one can get the solution as a function
of $a$ and $\theta$. In Fig.~\ref{fig_ahmadjon} dependence of the
critical value of the electric charge $Q$ and radius of the
static limit surface $r$ which is corresponding to critical state on the
rotation parameter $a$ have been shown for several values of
$\theta$. The shaded region in the $Q-a$ plot corresponds to the
regular black hole with the static limit. The unshaded region in the
$Q-a$ plot corresponds to the regular black hole without static limit.
The $r-a$ plot represents the dependence of the radius of the
static limit on the rotation parameter which corresponds to the border
of shaded-unshaded regions.
{In order to find the same critical value of the charge one may use the conditions
\begin{eqnarray}\label{002}
g_{rr}(r,a,\theta,Q)=0, & \partial_{r}g_{rr}(r,a,\theta,Q)=0.
\end{eqnarray}
 In Fig.~\ref{fig_ahmadjon2} dependence of the
critical value of the electric charge $Q$ and radius of the static
limit surface $r$ which is corresponding to the critical state on
the rotation parameter $a$ have been shown for several values of
$\theta$. The shaded region in the $Q-a$ plot corresponds to the
regular black hole with two (outer and inner) event horizons. The
unshaded region in the $Q-a$ plot corresponds to the regular black
hole without event horizon. The $r-a$ plot represents the
dependence of the radius of the static limit on the rotation
parameter which corresponds to the border of shaded-unshaded
regions.}
The regular black hole with the
rotation parameter $a\sim1$ can have the horizon in the poles of
the black hole ($\theta=0,\pi$) even in the case when the value of
the charge $Q$ is very small. One may conclude that in the
presence of the rotation parameter the small  value of electric
charge may cause the elimination of the singularity.

{The critical values of the electric charge $Q_{\rm
cr}$ are different for the static limit and the event horizon. The
critical value of the electric charge for event horizon is more
rapidly decreasing with the increase of the rotation parameter $a$
with compare to that for the static limit. This means that the
event horizon disappears earlier with the increase of the electric
charge for the fixed value of the rotation parameter $a$. }

One can see from Fig.~\ref{fig_ahmadjon} and \ref{fig_ahmadjon2} that when the value of
the electric charge $Q\leq0.633$ the static black hole ($a=0$) has
horizon. If $Q\leq0.605$ even extreme black hole can have the
horizon in the equatorial plane ($\theta=\pi/2$).

\begin{figure*}[t!.]
\begin{center}
\includegraphics[width=0.32\linewidth]{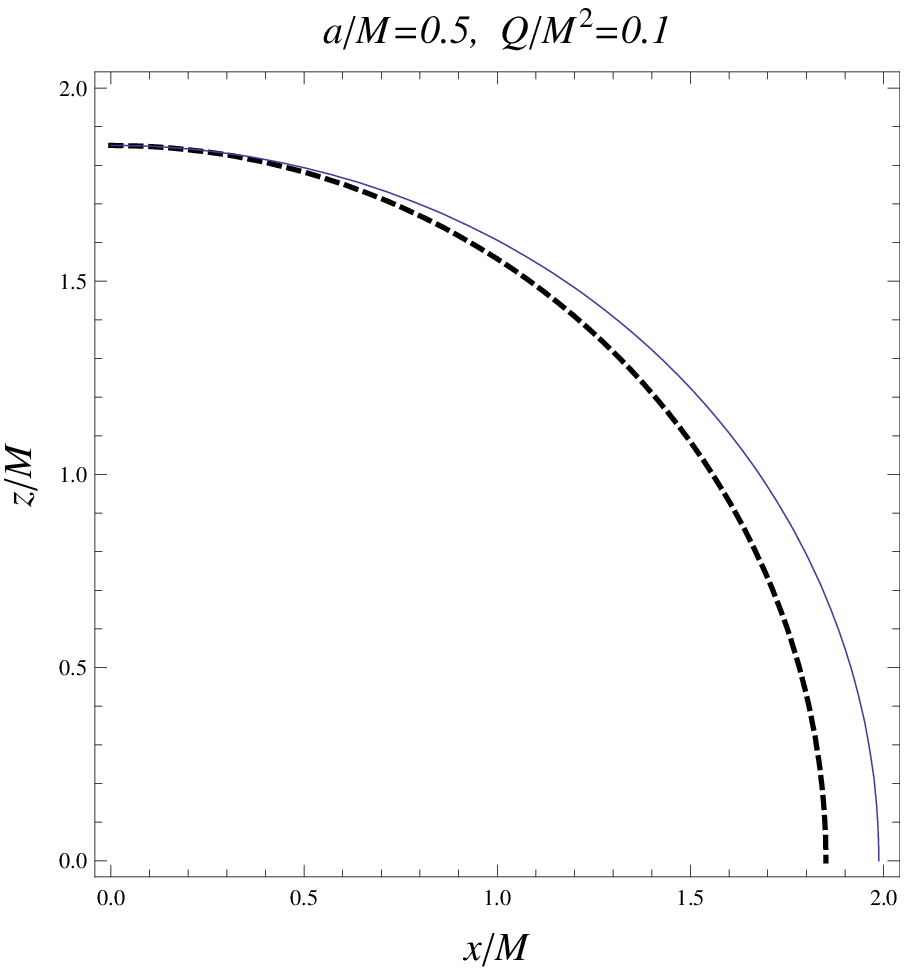}
\includegraphics[width=0.32\linewidth]{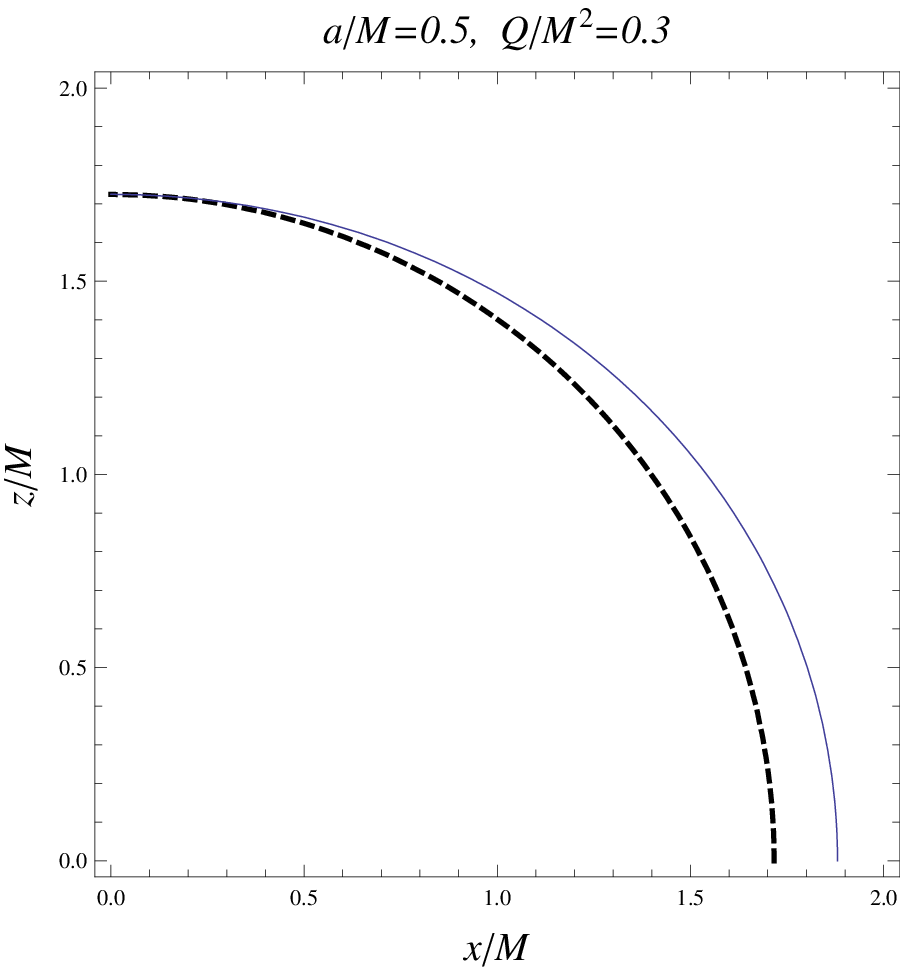}
\includegraphics[width=0.32\linewidth]{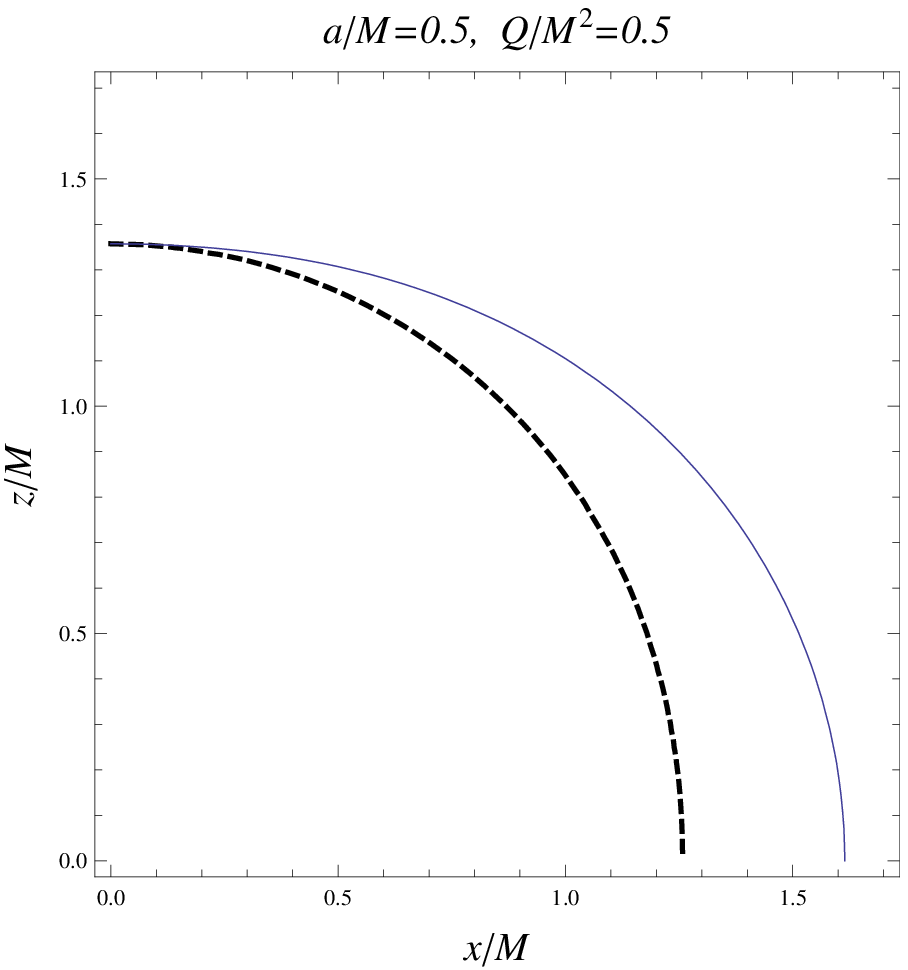}

\includegraphics[width=0.32\linewidth]{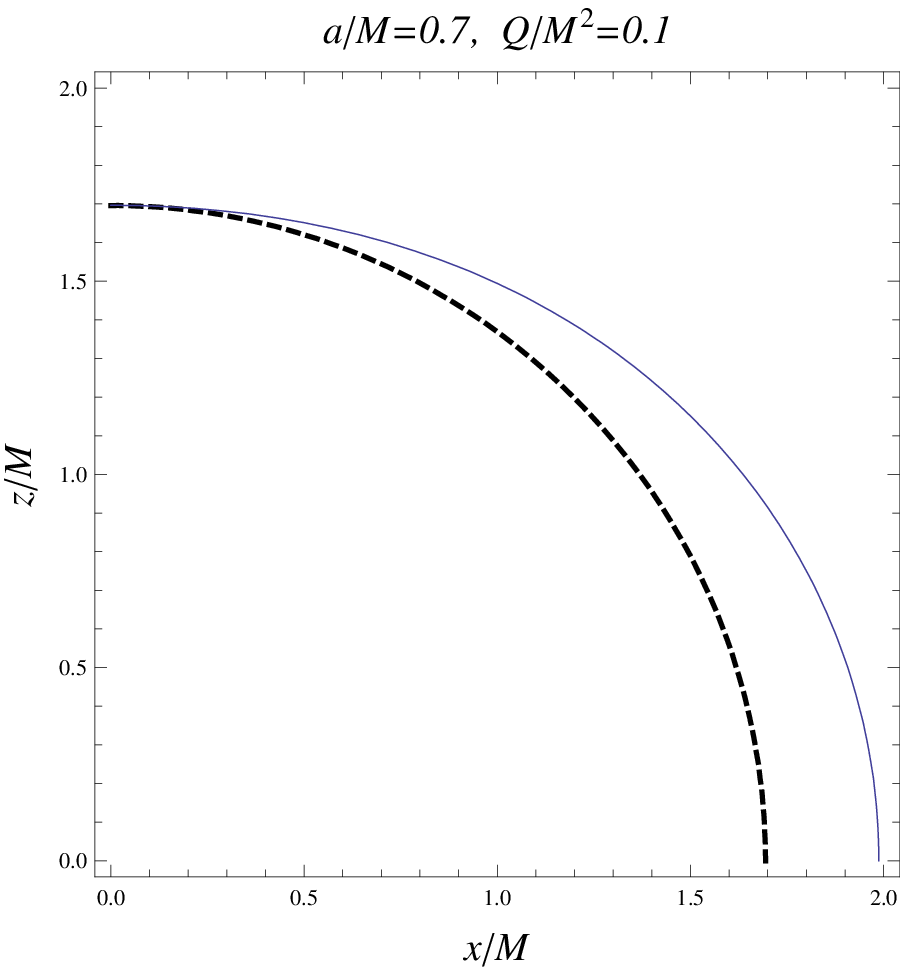}
\includegraphics[width=0.32\linewidth]{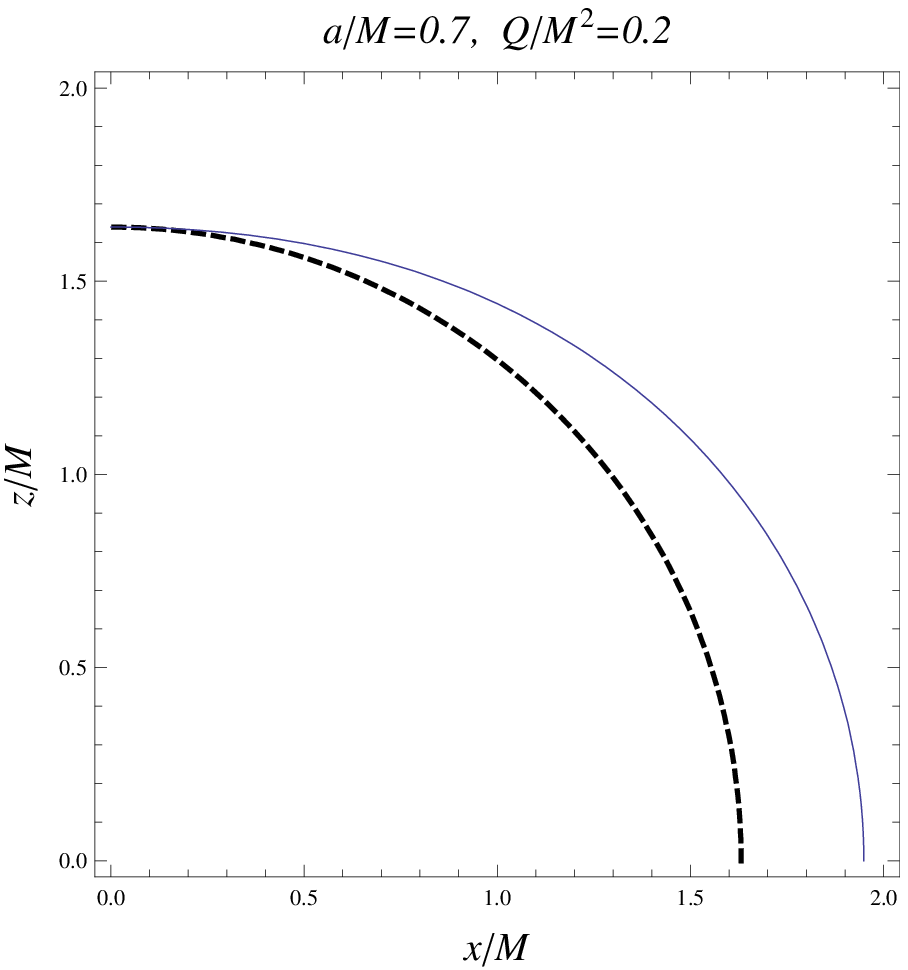}
\includegraphics[width=0.32\linewidth]{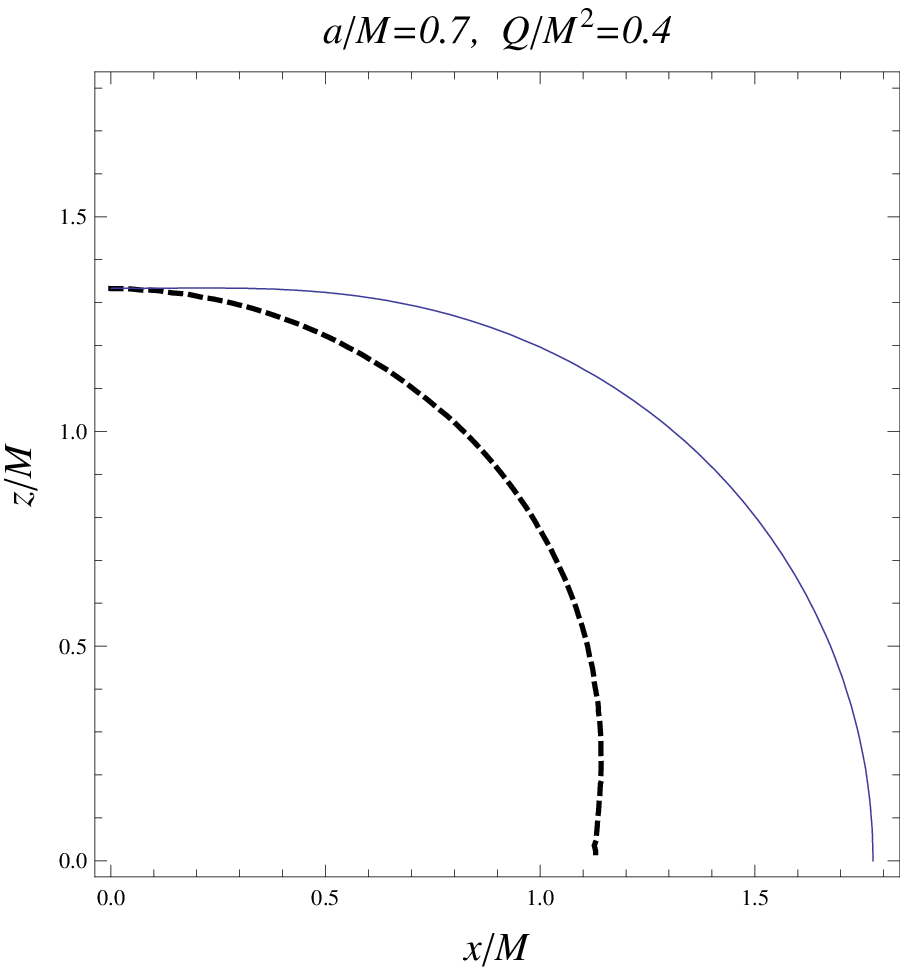}

\includegraphics[width=0.32\linewidth]{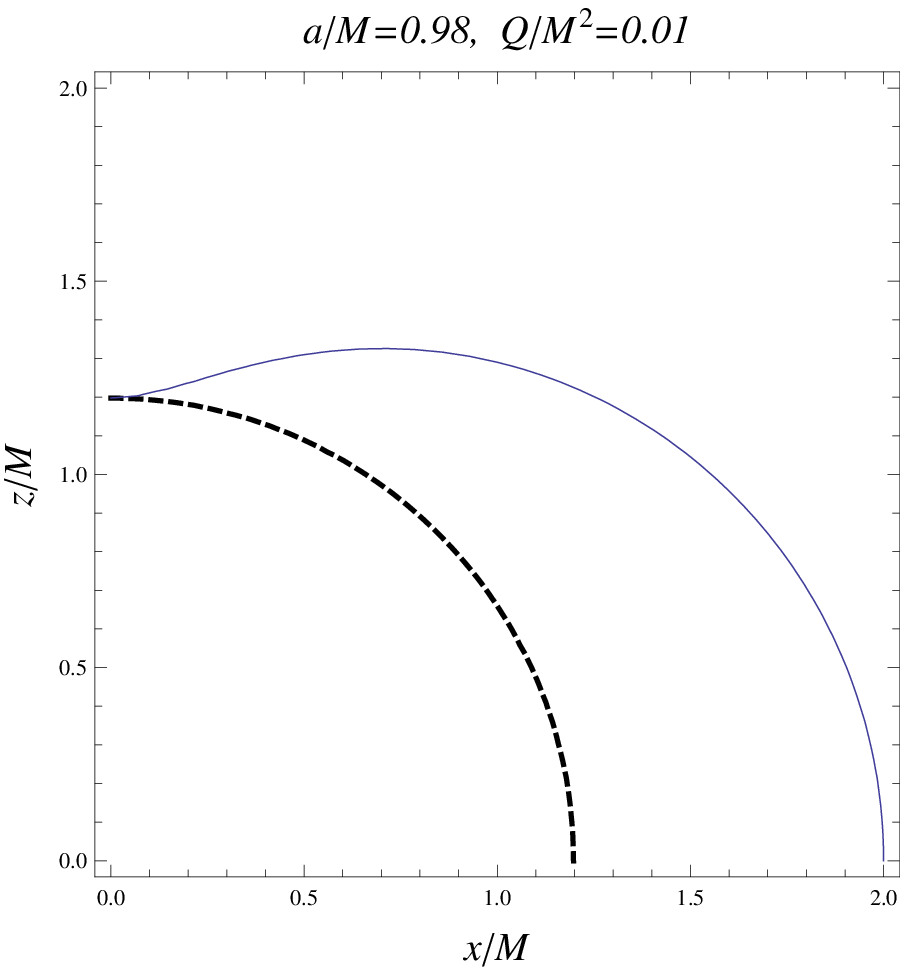}
\includegraphics[width=0.32\linewidth]{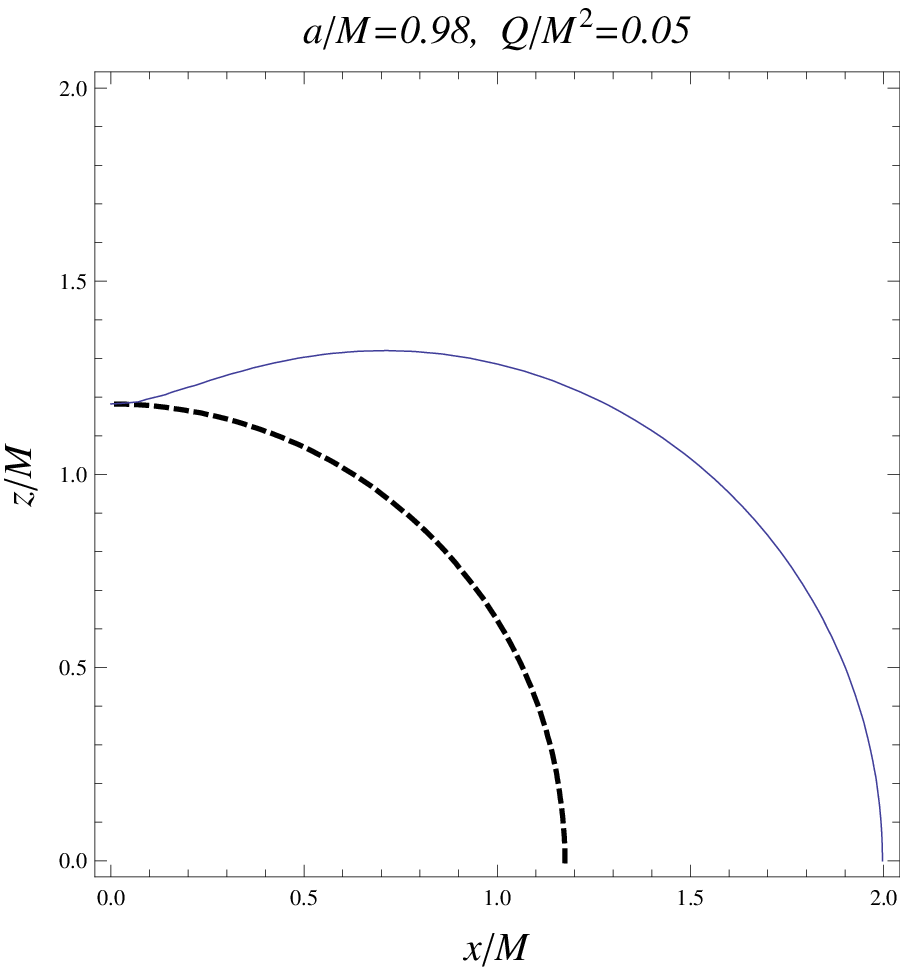}
\includegraphics[width=0.32\linewidth]{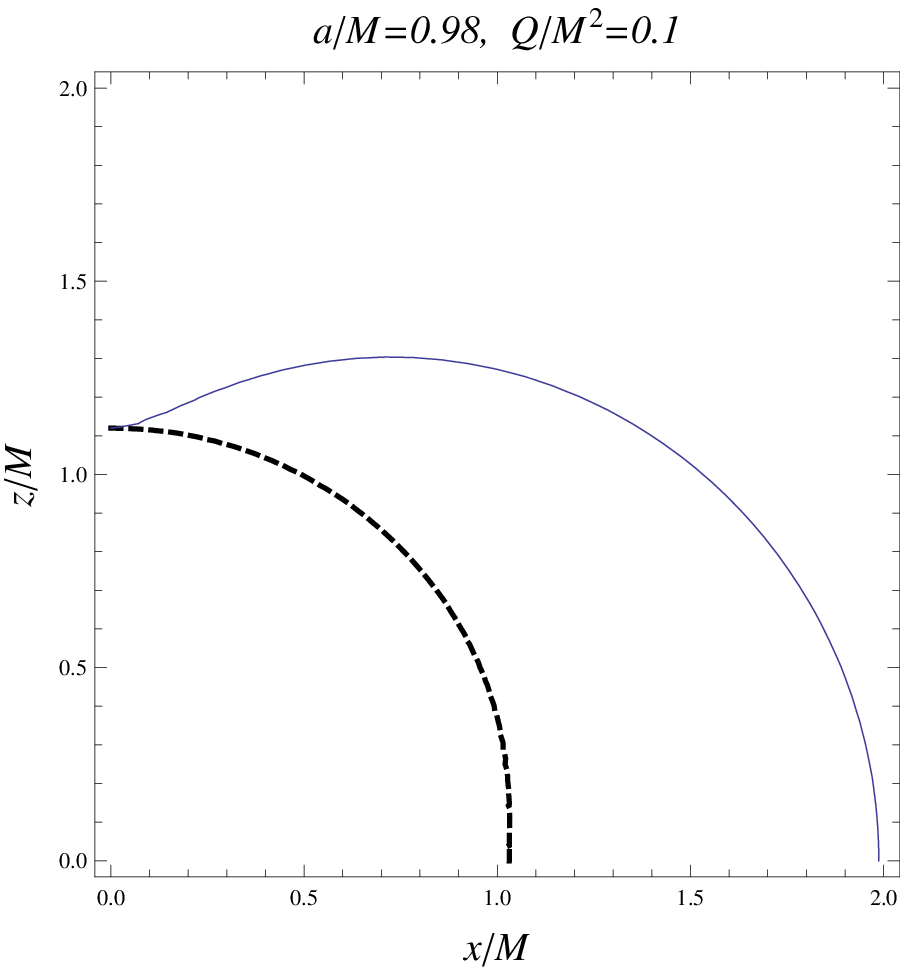}

\end{center}
\caption{\label{fig_ahmadjon3} The shape and size of the
ergosphere for the different values of the rotation parameter $a$
and electric charge $Q$.}
\end{figure*}

{Fig.~\ref{fig_ahmadjon3} provides the shape and
size of the ergoregion  in the $x-z$ plane where $z=r\cos\theta$
and $z=r\sin\theta$. With the increase of the electric charge one
can observe the increase of the relative shape and size of the
ergosphere. Note that for the values of electric charge with
$Q\geq Q_{\rm cr}$ the event horizon and static limit both
disappear. }

As the further 
step we  study 
the question of
satisfying the weak 
energy condition and 
choose the 
Locally non
rotating frame 
(LNRF) in order 
to get the stress-energy 
tensor in
diagonal form, 
namely, 
$T^{\alpha \beta}=(\rho,\ P_1,\  P_2,\
P_3)$. 
Then the weak 
energy condition 
reads as \cite{Bambi}
$\rho\geq0, \ \ \rho+P_i\geq0$, where $i=1,\ 2,\ 3$.

Finally, we express the spacetime geometry in
frame of the LNRF and study the behavior of the angular velocity
of these frames. The orthonormal tetrad of the LNRF has the
following form
\begin{eqnarray}
\omega^{t}&=&\left|g_{tt}-g_{\phi\phi}\Omega^2_{_{\rm LNRF}}\right|^{1/2} dt \ ,\\
\omega^{r}&=&\left|g_{rr}\right|^{1/2} dr \ ,\\
\omega^{\theta}&=&\left|g_{\theta\theta}\right|^{1/2} d\theta \ ,\\
\omega^{\phi}&=&\left|g_{\phi\phi}\right|^{1/2}d\phi-\left|g_{\phi\phi}\right|^{1/2}\Omega_{_{\rm LNRF}} dt \ ,
\end{eqnarray}
where
\begin{eqnarray}
\Omega_{_{\rm LNRF}} =\frac{a
(1-\tilde{f}(r,\theta))}{\Sigma-a^2(\tilde{f}(r,\theta)-2)\sin^2\theta}
\end{eqnarray}
is the angular velocity of the LNRF frame.

\begin{figure*}[t!.]
\begin{center}
\includegraphics[width=0.45\linewidth]{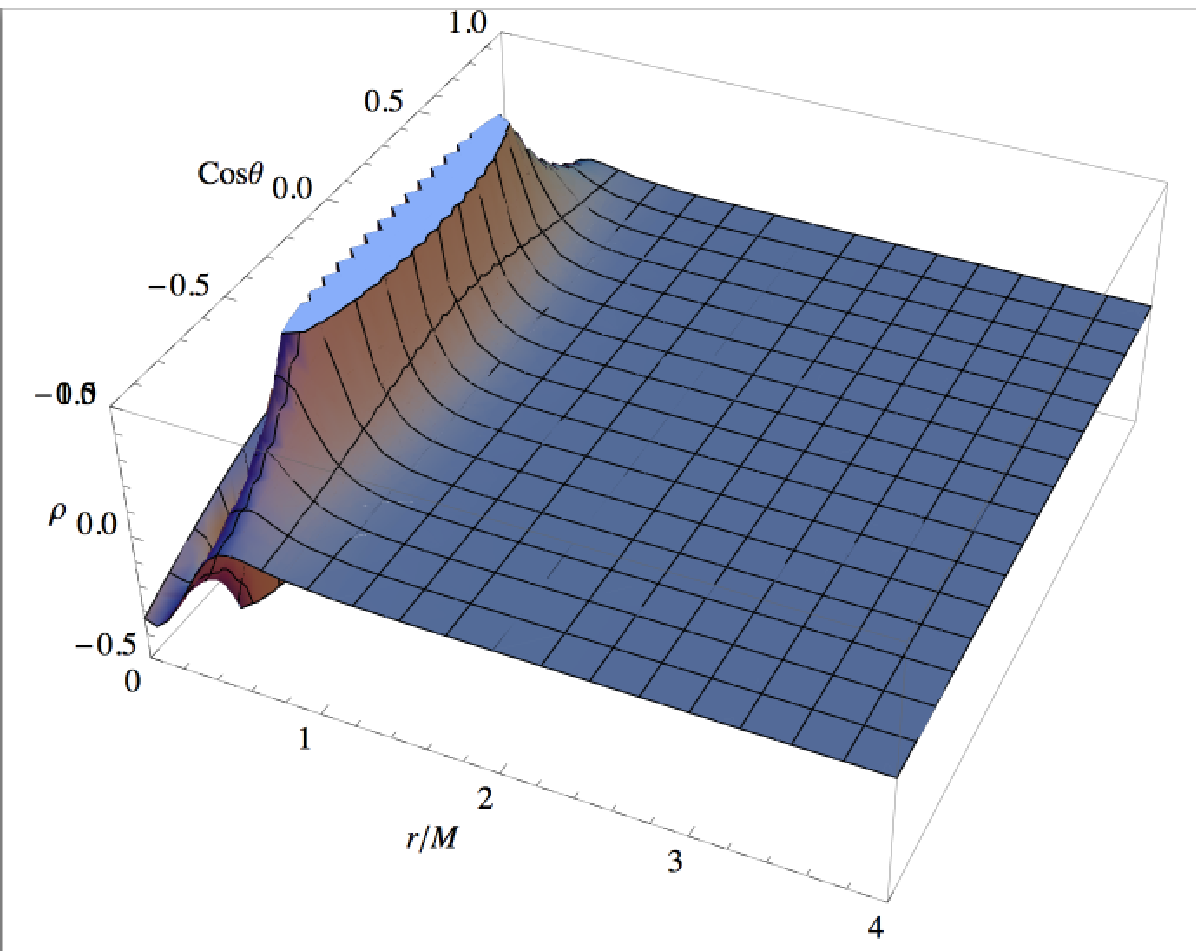}
\includegraphics[width=0.45\linewidth]{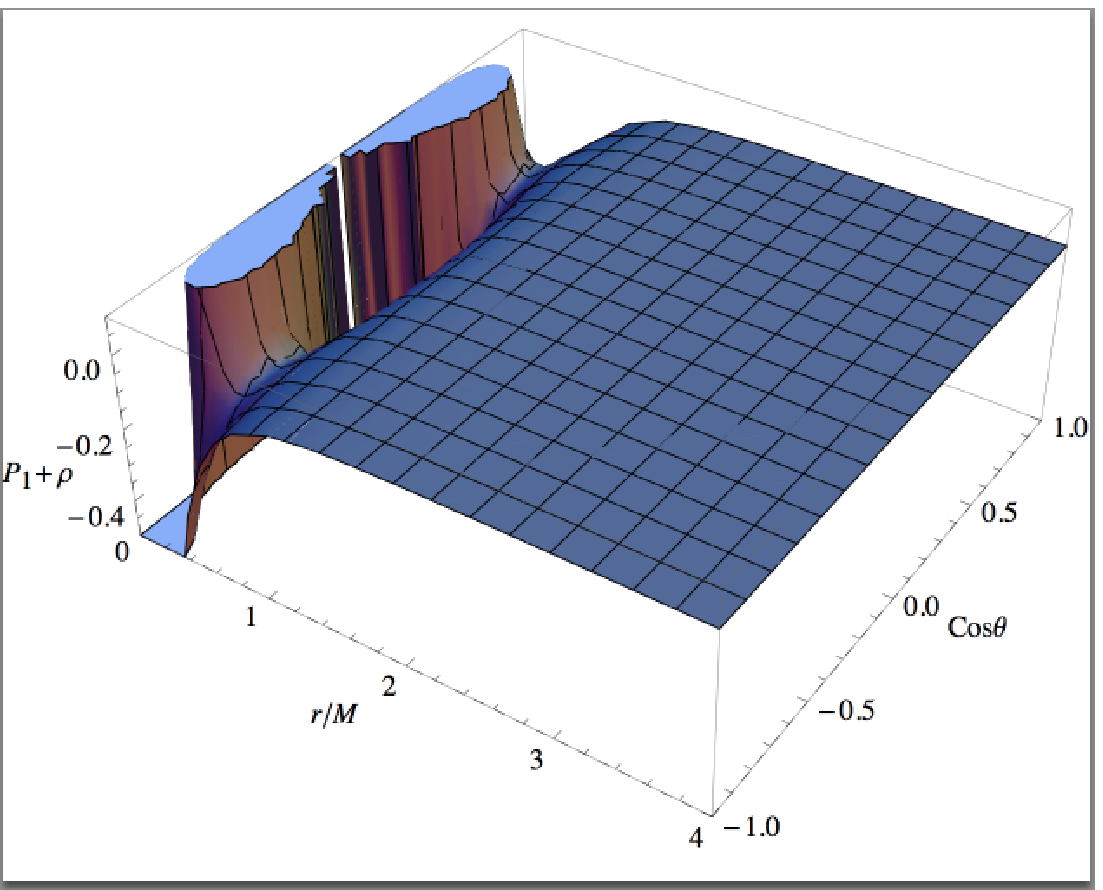}

\includegraphics[width=0.45\linewidth]{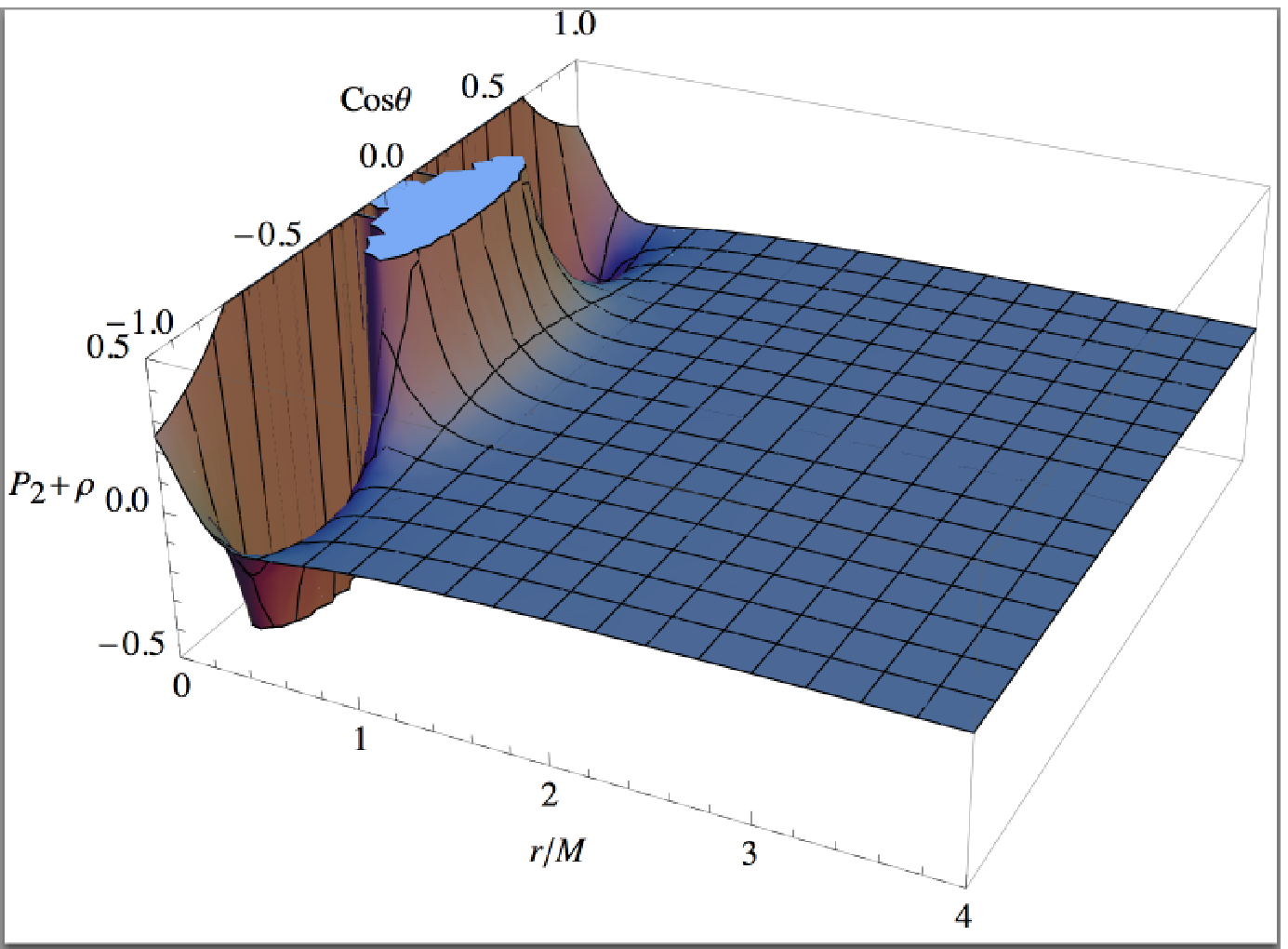}
\includegraphics[width=0.45\linewidth]{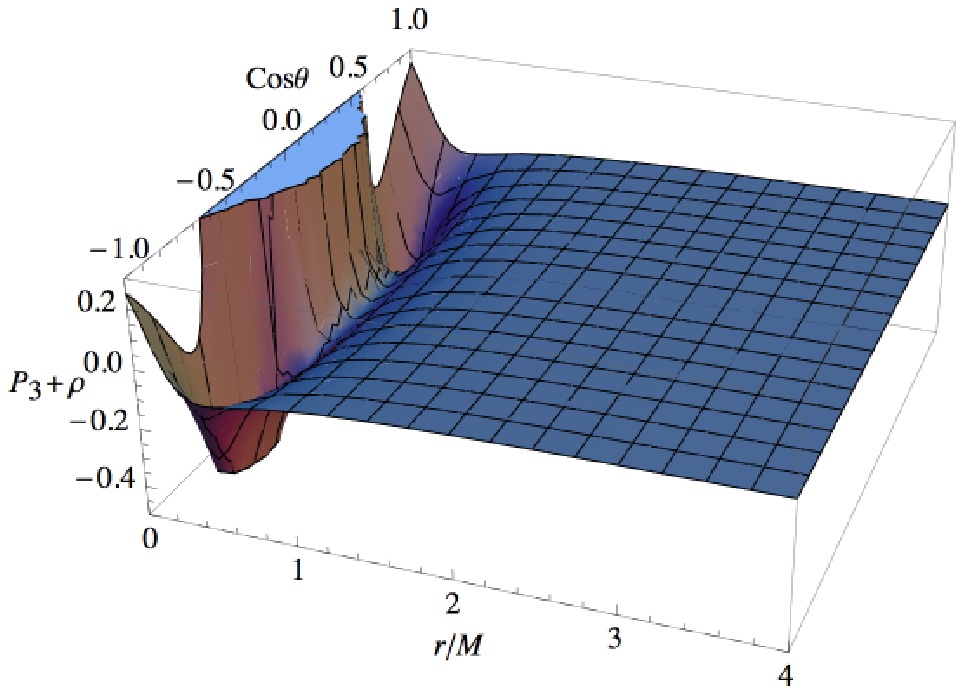}

\end{center}
\caption{\label{fig_ahmadjon4} The radial and angular dependence
of $\rho$, $P_1+\rho$, $P_2+\rho$, and $P_3+\rho$ for the given
value of the rotation parameter $a/M=0.5$ and electric charge
$Q/M=0.9$.}
\end{figure*}

In Fig.~\ref{fig_ahmadjon4} the  3D plot of the
radial and angular dependence of density and pressures is shown
for the given value of the rotation parameter $a/M=0.5$ and electric charge
$Q/M=0.9$. One can
easily see that the weak energy condition is violated near the nonsingular origin of rotating regular black hole. 

%
%  conclusion
%
%

\section{Conclusion}

In this paper, we used a regular black hole solution with the
source with the nature of nonlinear electrodynamics obtained by
Ay\'{o}n-Beato and Garc\'{i}a~\cite{Eloy1,Eloy,Eloy2} to generate
rotating regular black hole solution which includes the
Ay\'{o}n-Beato-Garc\'{i}a and  Kerr metrics as special cases.
Considered Newman-Janis algorithm uses static solution to generate
rotating solutions without touching the field equation and very
useful in order to get rotating black hole solutions.

The relation between Einstein vacuum solution and any non vacuum
solution of general relativity opens new direction in studying the
properties of the new solution with nonlinear electrodynamic
source. Obviously, when electric charge is vanishing, the solution
reduces to the vacuum one. Here we have obtained an exact rotating
regular BH solution in the framework of general relativity.
Obtained solution gives opportunity to study the geometrical and
causal structures, as well as test particles motion around
rotating regular BH, which will be subject of the future projects.
On the other hand, it could be also very
interesting to compare the  rotating Ayon-Beato-Garcia regular
spacetimes without horizons to the Kerr naked singularity
spacetimes, testing if the interesting and unusual physical
phenomena occuring in the Kerr naked singularity
spacetimes~\cite{ss10}  could arise in the regular rotating
spacetimes too.

\begin{acknowledgments}

The~authors acknowledge the~project Supporting
Integration with the~International Theoretical and Observational
Research Network in Relativistic Astrophysics of Compact Objects,
CZ.1.07/2.3.00/20.0071, supported by Operational Programme
\emph{Education for Competitiveness} funded by Structural Funds of
the~European Union. One of the authors (ZS) acknowledges the
Albert Einstein Center for gravitation and astrophysics supported
by the~Czech Science Foundation No.~14-37086G. A.A. and B.A.
thank the Goethe University, Frankfurt am Main, Germany and
Faculty of Philosophy and Science, Silesian University in Opava
(Czech Republic) for the warm hospitality. This research is
supported in part by Projects No. F2-FA-F113, No. EF2-FA-0-12477,
and No. F2-FA-F029 of the UzAS and by the ICTP through the
OEA-PRJ-29 and the OEA-NET-76 projects and by the Volkswagen
Stiftung (Grant No. 86 866).

\end{acknowledgments}

\label{lastpage}

\end{document}